\documentclass[12pt]{iopart}
 \input epsf.sty
\begin{document}

%%Useful symbols%%%%%%%%%%%%%%%%%%%%%%%%%%%%%%%%%%
\def\CA{{\cal A}}
\def\CB{{\cal B}}
\def\CC{{\cal C}}
\def\CD{{\cal D}}
\def\CE{{\cal E}}
\def\CF{{\cal F}}
\def\CG{{\cal G}}
\def\CH{{\cal H}}
\def\CI{{\cal I}}
\def\CJ{{\cal J}}
\def\CK{{\cal K}}
\def\CL{{\cal L}}
\def\CM{{\cal M}}
\def\CN{{\cal N}}
\def\CO{{\cal O}}
\def\CP{{\cal P}}
\def\CQ{{\cal Q}}
\def\CR{{\cal R}}
\def\CS{{\cal S}}
\def\CT{{\cal T}}
\def\CU{{\cal U}}
\def\CV{{\cal V}}
\def\CW{{\cal W}}
\def\CX{{\cal X}}
\def\CY{{\cal Y}}
\def\CZ{{\cal Z}}

%macros
\newcommand{\todo}[1]{{\em \small {#1}}\marginpar{$\Longleftarrow$}}
\newcommand{\labell}[1]{\label{#1}}
\newcommand{\bbibitem}[1]{\bibitem{#1}}
\newcommand{\llabel}[1]{\label{#1}\marginpar{#1}}

% macros for the conical defect paper
\newcommand{\sphere}[0]{{\rm S}^3}
\newcommand{\su}[0]{{\rm SU(2)}}
\newcommand{\so}[0]{{\rm SO(4)}}
\newcommand{\bK}[0]{{\bf K}}
\newcommand{\bL}[0]{{\bf L}}
\newcommand{\bR}[0]{{\bf R}}
\newcommand{\tK}[0]{\tilde{K}}
\newcommand{\tL}[0]{\bar{L}}
\newcommand{\tR}[0]{\tilde{R}}

\newcommand{\btzm}[0]{BTZ$_{\rm M}$}
\newcommand{\ads}[1]{{\rm AdS}_{#1}}
\newcommand{\eAds}[1]{{\rm EAdS}_{#1}}
\newcommand{\ds}[1]{{\rm dS}_{#1}}
\newcommand{\eds}[1]{{\rm EdS}_{#1}}
\newcommand{\sph}[1]{{\rm S}^{#1}}
\newcommand{\gn}[0]{G_N}
\newcommand{\SL}[0]{{\rm SL}(2,R)}
\newcommand{\cosm}[0]{R}
\newcommand{\hdim}[0]{\bar{h}}
\newcommand{\bw}[0]{\bar{w}}
\newcommand{\bz}[0]{\bar{z}}
\newcommand{\be}{\begin{equation}}
\newcommand{\ee}{\end{equation}}
\newcommand{\bea}{\begin{eqnarray}}
\newcommand{\eea}{\end{eqnarray}}
\newcommand{\pat}{\partial}
\newcommand{\lp}{\lambda_+}
\newcommand{\bx}{ {\bf x}}
\newcommand{\bk}{{\bf k}}
\newcommand{\bb}{{\bf b}}
\newcommand{\BB}{{\bf B}}
\newcommand{\tp}{\tilde{\phi}}
\hyphenation{Min-kow-ski}

%%Commonly used constants and symbols%%%%%%%%%%%%%%%%%%%%%%%%%
\def\apr{\alpha'}
\def\str{{str}}
\def\lstr{\ell_\str}
\def\gstr{g_\str}
\def\Mstr{M_\str}
\def\lpl{\ell_{pl}}
\def\Mpl{M_{pl}}
\def\varep{\varepsilon}
\def\del{\nabla}
\def\grad{\nabla}
\def\tr{\hbox{tr}}
\def\perp{\bot}
\def\half{\frac{1}{2}}
\def\p{\partial}
\def\perp{\bot}
\def\eps{\epsilon}

%%%%%%%%%%%%%%%%%%%%%%%%%%%%%%%%%%%
% Erich's macros:
\renewcommand{\thepage}{\arabic{page}}
\setcounter{page}{1}

\rightline{hep-th/0404075, UPR-T-1072}
%\title[Time-dependent Universes and String Theory]{Time-dependent Universes in String Theory}
\title[Accelerating Universes and String Theory]{Accelerating Universes and String Theory}

\author{Vijay Balasubramanian}

\address{David Rittenhouse Laboratories, The University of Pennsylvania, Philadelphia, PA 19104, USA}

\begin{abstract}
This article reviews  recent developments in the study of universes with a positive cosmological constant in string theory.
%of expanding and time-dependent universes in string theory.   Four areas are discussed: (i) Universes 
%with a positive cosmological constant, (ii) Holography and time-dependence, (iii) Simple solvable 
%models, (iv) Decaying branes.
\end{abstract}

%\newpage

\section{The Challenge of Time}
\label{intro}

It is evident that the world around us changes in time.   At cosmological scales there is strong evidence that the universe was much hotter and denser some 13 billion years ago and has expanded and cooled since then, probably even undergoing an early period of exponentially accelerating expansion \cite{WMAP}.  Apparently, even today a significant component of the matter in the universe is ``dark energy" \cite{WMAP,supernovae}, which might be a cosmological constant or a rolling scalar field \cite{quint} or something else, that is causing an accelerating expansion of spacetime.   Despite these important motivations, very little is known about the basic features of string theory in expanding and time-dependent spacetimes, about the construction of realistic expanding universes  from fundamental theory, about the nature of ``dark energy''  and the cosmological constant problem, and, most fundamentally, about ``what time is".     More concretely, can the many light scalar fields (moduli) that arise in string compactifications acquire potentials that lead to slow roll inflation or quintessence, or perhaps can they be stabilized at a positive potential minimum giving a small cosmological constant?  Expanding universes often begin in a big bang -- can such spacelike curvature singularities including the ones inside Schwarzschild black holes be resolved in string theory?  Is there a holographic description of universes with a positive cosmological constant akin to the ones available for a negative cosmological constant?    Is information lost in black holes?

It is important to mention that a fundamental challenge for string theory is that even the basic perturbative formulation of the theory is not understood in a general time-dependent spacetime.   Recall that the sigma model approach to string perturbation theory constructs the partition function as a functional of the spacetime background 
\begin{equation}
Z[G_{\mu\nu},\cdots] = \int {\cal D} g_{ij} \, {\cal D}X^\mu e^{ i \int \sqrt{g}\,   G_{\mu\nu}(X) \, \partial_i X^\mu \partial_j X^\nu g^{ij} \ +  \ \cdots }
\labell{stringpert}
\end{equation}
as a sum over worldsheet metrics ($g_{ij}$) and embeddings ($X^{\mu}$) in a fixed spacetime background.   This object is usually defined by continuing both the worldsheet and the spacetime to Euclidean signature.   However,  a general time-dependent spacetime does not have a Euclidean section attained simply by taking $t \to it$; such continuations generally produce a complex metric.   Continuing the worldsheet and not the spacetime is not well-defined since it leads to an action that is not bounded from below.  (Also, since the saddlepoints of (\ref{stringpert}) arise from classical motions of the string, a Euclidean worldsheet propagating in a Lorentzian spacetime will not have a good saddlepoint.)     Simply integrating over Lorentzian worldsheet metrics is also problematic -- even the simplest interaction of string theory, the pants diagram, does not have an everywhere non-singular Lorentzian metric.   If we should integrate over singular metrics on Lorentzian worldsheets, the rules for doing so have not been defined.  Despite this essential difficulty we can try to make progress by learning general lessons from spacetimes which {\it can} be Wick rotated to Euclidean signature and also by analyzing solvable toy examples where this cannot be done.

Since the primary motivation for exploring the issues described above is the observed acceleration of the universe, this article is a rapid snapshot of tools, techniques and recent developments in the study of universes with a positive cosmological constant within string theory.

\section{Universes with a positive cosmological constant}
\label{sec:lambda}

Our main motivation for interest in time-dependent universes is cosmological.  Big bang and inflationary cosmologies change in time, and the apparent observation of $\Lambda > 0$ today implies an accelerating expansion.   (See \cite{padmanabhan1} for a recent review of the physics of the cosmological constant.)  The equations of motion with a positive cosmological constant are derived from the action
\begin{equation}
S =  - {1 \over 16 \pi G_N} \int_{{\cal M}} d^{d+1}x \, \sqrt{-g} \, (R(g) \, + \, 2  \Lambda) ~+~ {1 \over 8\pi G_N} \, \int_{{\cal I}^+}^{ {\cal I}^+} \sqrt{h}\,  d^{d}x \,  K
\labell{dsaction1}
\end{equation}
where $2\Lambda = d(d-1)/\ell^2$  is the cosmological constant, ${\cal M}$ is the bulk manifold, $h$ is the induced metric on the early and late time boundaries ${\cal I}^\pm$, and $K$ is the trace of the extrinsic curvature of the boundaries.  
The maximally symmetric solution (with  $(d(d+1)/2)$ isometries in $d+1$ dimensions) is de Sitter (dS) space which can be written as the hyperboloid
\begin{equation}
 -(X^0)^2 + (X^1)^2 +(X^2)^2 + \cdots (X^{d+1})^2 = \ell^2
 \labell{hyperboloid1}
 \end{equation}
 embedded in ${\bf R}^{1,d+1}$. Notice that if $\Lambda \sim 1/\ell^2 < 0$, these are exactly the defining equations of Euclidean anti de-Sitter (EAdS) space about which a great deal is known in string theory.   Unlike AdS, de Sitter space, with a positive vacuum energy, cannot be supersymmetric.  A rough reason for this is easily seen from the form of the scalar potential in $N=1$ supergravity:
 \begin{equation}
 V = e^K \, \sum_{ij} {\cal G}^{ij}\,  D_i W \, D_j W - 3W^2
 \end{equation}
 where $K, W, {\cal G}$ are the $N=1$ Kahler potential, superpotential, and metric on the space of scalars $\phi_i$.   Supersymmetry obtains when $D_i W = 0$, which can only lead to $V \leq 0$. (Formal de Sitter superalgebras were worked out in \cite{dsalgebra}, but they have non-compact R-symmetry groups, which lead to ghostlike fields in the Lagrangian (wrong sign kinetic terms), and  have no non-trivial representations  on  a positive Hilbert space.)
 
 The  metric of de Sitter space can be written as
 \begin{equation}
 ds^2 = -dt^2 + \cosh^2(t/\ell) \, d\Omega_{d}^2
 \labell{dsglobal}
  \end{equation}
 which describes a sphere that contracts from infinite size  and then grows exponentially again.   The spacetime boundaries ${\cal I}^{\pm}$ at $t \to \pm\infty$ are spheres.  Interestingly, taking $t \to it$ gives a Euclidean section which is precisely a round $d+1$ sphere:
 \begin{equation}
 ds_E^2 =  \ell^2( d\tau^2 + cos^2(\tau) \, d \Omega_d^2)
 \labell{dseuclid1}
 \end{equation}
Two other forms of the metric are 
 \begin{eqnarray}
 ds_I^2 &=& -dt_I^2 \,  + \, e^{2t_I/\ell}    \, d\vec{x}^2
 \labell{dsinflat} \\
 ds_{II}^2 &=& - (1 - {r^2 / \ell^2}) \, dt_{II}^2 \,  + \,  (1 - {r^2 / \ell^2})  \, dr^2 + \,  r^2 \, d\Omega_{d-1}^2
 \labell{dsstatic}
 \end{eqnarray}
 The first coordinates, marked $I$ in the Penrose diagram Fig.~1B, cover the ``inflating patch'' of de Sitter space.   This section of de Sitter space is patched in after a Big Bang to describe the inflating era.   The second coordinates, marked $II$ in Fig.~1B, cover the ``static patch'', describing the universe as seen by an inertial observer in dS located at $r=0$.   The dotted lines in Fig.~1B are the cosmological horizons seen by such an observer.   The top and bottom boundaries of the figure mark the future and past conformal boundaries of de Sitter space, reached as global $t \to \pm \infty$.     Looking at the Penrose diagram one of the strange features of de Sitter space is apparent -- two observers on a sufficiently late time surface (a horizontal line running across  Fig.~1B) will never be able to communicate.  This immediately implies that the standard relation between correlation functions and an S-matrix that we rely on in field theory and perturbative string theory must break down.  In fact, this is a difficulty for many accelerating universes including typical quintessence models (\cite{quintprob}).
 
 During inflation the  the universe is modelled as expanding in a de Sitter phase with a metric of the form (\ref{dsinflat}).  Close to the Big Bang, this metric is matched onto some other geometry or to some non-geometric phase of the universe, while after the exit from inflation, radiation and matter dominated FRW universes take over.  The exponential expansion in (\ref{dsinflat}) makes inflation a natural cosmic accelerator.   The $\sim$60 e-foldings that occur during inflation have the potential to imprint physics far above the Planck scale in the CMBR spectrum \cite{brandenberger}.    A number of groups have considered how stringy modifications to physics at and above the Planck scale, such as modified commutation relations or non-locality, could give rise to observable effects in the next generation of CMBR experiments \cite{brandenberger,greene,ulf,burgess,mersini}.   Whether the effects are observable depends in part on what one considers an acceptable initial condition for inflation \cite{kaloper}.   Specifically, conventional inflation begins in the standard adiabatic vacuum state for de Sitter space.  However, it is possible to choose the initial state from a 1-parameter family of de Sitter invariant vacua (called $\alpha$-vacua) \cite{ancient}.  (See \cite{bms,alphons} for recent discussions of these vacua and further references.)    While these vacua have some undesirable properties (for example, they have non-local correlations that make it challenging to consistently define loop diagrams \cite{finnds,loweds,collins}) one can consider approximate $\alpha$ vacua that approach the adiabatic vacuum at sufficiently high energies.   For example, \cite{ulf} selects such initial states by demanding that, as modes inflate to come above the Planck scale, they should be in their vacuum rather than being thermally excited as they would usually be.   The physics of $\alpha$-states was extensively discussed in \cite{alphons} along with a discussion of imprints in the CMBR and the possibility of making the vacuum selection parameter $\alpha$ a dynamical variable.   While it is still not conclusively argued that cosmic signatures of transplanckian physics will be available, this possibility remains an intriguing avenue for exploration.

  Fig.~1C shows  dS and Euclidean AdS as different hyperbolic sheets of revolution in ${\bf R}^{1,d}$.    This makes it apparent that the two spaces share the same boundary when they are seen as real sections of the same complexified manifold  -- the surfaces ${\cal I}^\pm$ in de Sitter space are the same spheres $S^d$ as the boundaries of Euclidean AdS.  It is also immediately apparent that the asymptotic symmetry group of $\ds{d+1}$ will be the same as that of $\eAds{d+1}$, namely the Euclidean conformal group in d dimensions, $SO(d+1,1)$.

\begin{figure}
  \begin{center}
 \epsfysize=2in
   \mbox{\epsfbox{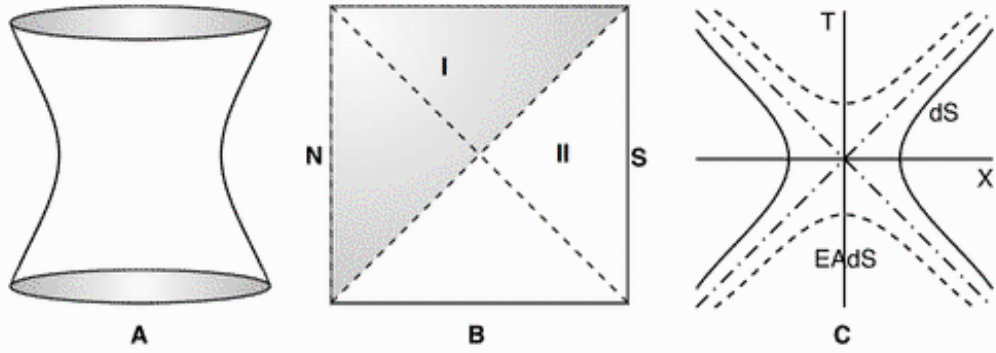}}
    \caption{(A) Global de Sitter space is a hyperboloid (see eq. \ref{dsglobal}).  Time runs up and equal time sections are spheres.   (B) Penrose diagram of de Sittter space.   Horizons are dashed lines, Regions I (shaded) and II are the inflating and static patches respectively.  The left boundary and the right boundary (marked N and S) are the north and south poles of the equal time sections of the global spacetime.  The Euclidean surfaces at the top and bottom boundaries are future and past infinity.  (C) de Sitter (dS) and Euclidean anti de Sitter (EAdS) are hyperboloids of revolution around the vertical axis.  The two sheets with $|T|>|X|$ in the figure are both EAdS, while the lines with $|X| > |T|$ are part of a single hyperboloid of revolution around the $T$ axis that makes up global de Sitter as in (A).  Note that EAdS and dS share the same asymptotic boundary structure as $|T|,|X| \to \infty$ and hence will share the same asymptotic isometry group.
 }
 \end{center}
\label{fig:desitter}
\end{figure}

\section{Thinking about de Sitter holography}

The relation between Euclidean AdS and Lorentzian de Sitter space, and in particular the shared asymptotic structure suggests that it might be possible to give a holographic dual to de Sitter in terms of a Euclidean conformal field theory \cite{stromdscft,wittends,BdBM1,stromds2,BdBM2}.   A basic argument against such a possibility arises from the entropy of de Sitter space.    

As seen by an inertial observer (\ref{dsstatic}), de Sitter space has a cosmological horizon.  Associating an entropy with this area leads to
\begin{equation}
S = {A \over 4 G_N} = { (2\ell)^{d-1} \pi^{d/2} \over G_N \Gamma(d/2)}
\labell{dsentropy}
\end{equation}
 The introduction of matter into de Sitter space causes the cosmological horizon to {\it contract} towards the inertial observer with the net effect that total entropy (matter + horizons), rather unusually, is {\it reduced} by the introduction of matter (including black holes) \cite{boussoentropy1}:
 \begin{equation}
 S_{{\rm matter}} + S_{{\rm black~holes}} + S_{{\rm Cosm.~Horizon}} \leq S_{{\rm de~Sitter}}
 \labell{dsentbound}
 \end{equation}
 Various arguments for and interpretations of this entropy are discussed in \cite{BHM}.    In particular, given that a round sphere is Euclidean de Sitter space (\ref{dseuclid1}), we could regard the sphere in $\ads{d} \times S^p$ compactifications as Euclidean de Sitter space and estimate the number of degrees of freedom in the CFT dual to AdS that are required to describe dynamics within the sphere.  It is interesting that this number scales exactly like the de Sitter entropy (\ref{dsentropy}) \cite{BHM}.  For example,  regard the $\ads{5} \times S^5$ compactification of string theory (with scale $\ell$ and dual to $SU(N)$ gauge theory)  as $\ads{5} \times {\rm Euclidean}~dS^5$.  We would then associate an entropy
 \begin{equation}
 S \sim {\ell^3 \over G_5 }\sim N^2
 \end{equation}
 with the  ``de Sitter'' factor where $\ell$ is the scale of the sphere and $G_5$ is the 5d Newton constant. The last equation is obtained by translating from spacetime to dual field theory variables \cite{BHM}.   Interestingly, the dual $SU(N)$ field theory has $O(N^2)$ degrees of freedom in its matrices and these are indeed used to describe the dynamics of the sphere.

The gravitational entropy of black holes was the original inspiration for the holographic principle \cite{lennyhologram} which states that this entropy is a measure of the actual number of accessible states.   (For a review see \cite{boussoreview}.)  For a black hole we usually view the entropy as measuring the number of microstates that are consistent with the macroscopic charges.  But in a cosmological setting the interpretation is more problematic.   First of all, in a general expanding universe the classic Bekenstein-Hawking entropy bound has to be generalized to involve not the area bounding a volume at fixed time, but rather a lightsheet emanating from such a surface, in order to correctly account for the entropies of matter flowing into and out of a region \cite{covent}.     In fact, in the simple case of empty de Sitter space the formula (\ref{dsentropy}) suffices.  Nevertheless there is an interpretational problem because each inertial observer in de Sitter space sees a horizon to which he or she might assign an entropy.   How then should we interpret the entropy?  Is it a quantity measuring the physics accessible to a fixed observer, or does it refer to the universe as a whole?\footnote{See \cite{BHM} for many different possible interpretations.}

Notably, Fischler \cite{willydsent} and Banks \cite{banks00} have argued that we should interpret de Sitter entropy as indicating a finite dimensional Hilbert space of excitations for the entire universe, a situation that usually arises when a system is  described by a topological theory or if  there is a fundamental discreteness of variables.    This is a radical proposal -- indeed, neither field theory nor perturbative string theory would lead us to expect a finite dimensional Hilbert space.   This interpretation also requires a form of ``horizon complementarity" in which the degrees of freedom measured by a given inertial observer, and contributing to his or her measured entropy, must be ``complementary" or in some sense equivalent to the degrees of freedom seen by any other inertial observer.     It has been argued that black holes enjoy such a principle of complementarity \cite{complementarity}, but requiring it of cosmological horizons is an additional step.    Banks and Fischler have argued that  there is such principle of horizon complementarity and that it can help in defining the analog of an S-matrix to describe the observations made by an inertial observer in de Sitter space \cite{banksfisch}.    

In any case, if the dimension of the Hilbert space is indeed finite,  the spacetime cannot be dual to a CFT -- indeed, a single harmonic oscillator has an infinite dimensional Hilbert space.    Yet another argument \cite{recurrence}  states that if de Sitter space has a finite entropy as in (\ref{dsentropy}), we should expect Poincare recurrences in the dynamics seen by an inertial observer.   It is difficult to imagine how this would be reproduced in a dual CFT with an infinite number of degrees of freedom.

Despite these important objections, it is a productive exercise to examine what the structure of a dual to de Sitter would have to be like, given the known properties of de Sitter space \cite{stromdscft,wittends,BdBM1,stromds2,BdBM2}.  Besides elucidating various aspects of the symmetries and dynamics of de Sitter, thinking in this way has led to two important new results: (a) We obtain a definition of mass in an asymptotically de Sitter universe, a new result in classical gravity \cite{BdBM1}, (b) In an inflating universe, or one with a positive cosmological constant, one can productively organize equations describing flow in time as a sort of renormalization group flow.    The definition of mass also leads to a positive mass conjecture -- de Sitter space has the largest mass of any nonsingular spacetime with a given  cosmological constant.   Spaces with a larger mass are conjectured to develop singularities.  In this way, the finiteness of de Sitter entropy can be consistent with an infinite dimensional Hilbert space -- most states have too much mass and lead to singular evolution, and therefore will not be accounted for by the entropy of the horizons of non-singular spaces with a positive cosmological constant.

The sections below rapidly review various proposals for defining de Sitter holography and focus on the results (a) and (b) described above.

\subsection{Thinking about duality}

Given the relation with AdS described above, it is clear that the asymptotic symmetry group of de Sitter space will be the Euclidean conformal group.     This analogy suggests that the holographic dual to de Sitter space will involve a Euclidean CFT.  However, the AdS/CFT correspondence contains three other ingredients that are lacking in de Sittter space: (a) a dictionary mapping bulk fields in a specific stringy realization into boundary data, (b) an argument for correctness in terms of a decoupling limit on a stack of branes, and (c) a defining equation, 
\begin{equation}
Z_{\rm bulk}(\phi_0) = \int_{\phi \to \phi_0} {\cal D}\phi \, e^{iS(\phi)} =  \langle e^{\int_{\partial_{\cal M}} \phi_o {\cal O}}\rangle \, ,
\labell{adscftdef}
\end{equation}
 which relates the spacetime partition function as a functional of boundary data $\phi_0$ to the generating function of correlation functions of a dual.

Nevertheless, we can attempt proceed by analogy, and by using symmetry.    An important difference compared to AdS space is that de Sitter space has two conformal boundaries ${\cal I}^{\pm}$ which are in causal contact.    We might try to formulate the de Sitter analog of (\ref{adscftdef})
\begin{equation}
Z_{{\rm bulk}}(\phi^{{\rm in}}, \phi^{{\rm out}}) = \langle \exp({\int_{{\cal I}^-} \phi^{{\rm in}} {\cal O}^{{\rm in}} \ + \ \int_{{\cal I}^+} \phi^{{\rm out}} {\cal O}^{{\rm out}}} )\rangle \, ,
\end{equation}
but there are several challenges to doing this.  For example, de Sitter space does not enjoy the classic AdS division of mode solutions into ``normalizable" modes representing states in the bulk and in the dual, and ``non-normalizable modes" representing both boundary conditions for the bulk and sources in the dual field theory.\footnote{For related issues in formulating AdS holography in the presence of multiple boundaries see \cite{eternalBH,juanbtz,BKLT,eskoper,KOS,leviross,BNS,liatjuanmulti}
 which study spaces with horizons, without horizons and euclidean settings.}   Even more problematically the left hand side of this equation is a Lorentzian path integral and thus complex while the right hand side, defined on the Euclidean boundaries is naively real.

These subtleties led Strominger to propose that de Sitter space is dual to a CFT defined on one of the two de Sitter boundaries \cite{stromdscft}.   He suggested that the two boundaries be identified by an antipodal map, essentially since a light ray starting at a given point on the sphere at ${\cal I}^-$ reaches the antipodal point on the sphere at ${\cal I}^+$.   He proposed to then define the correlation functions of a CFT dual to de Sitter space by generalizing the perturbative formulae in \cite{BGL}.  Schematically the two-point function of the dual theory would be written in terms of a bulk-boundary propagator $K$ as
\begin{eqnarray}  
\langle O(\bb^1) O(\bb^2) \sim  lim_{t \to \infty} \int_{{\cal I}^-}   d\bb^1 \, d\bb^2 && \sqrt{\det{h_1}} \sqrt{\det{h_2}} \\  &&
\phi(\bb^1,t^1) \stackrel{\leftrightarrow}{\partial}_t    K(\bb^1,t^1;\bb^2,t^2)  \stackrel{\leftrightarrow}{\partial}_t \phi(\bb^2,t^2). \nonumber
\end{eqnarray}
where $\partial_t$ is a derivative in the direction normal to the boundary at $t \to \infty$, $K$ is a bulk-to-boundary propagator computed by rescaling the boundary limit of the Feynman propagator, $\bb^{1,2}$ are boundary points, and $h$ is the boundary metric.  The detailed expressions of course differ in the different de Sitter coordinate systems.  Also there is a question of whether we should associate operators with each of the two boundaries and how we deal with the fact that unlike AdS, all mode solutions are normalizable.  (See  \cite{spradvolds} for a  review of the proposal and extension to the different coordinate systems.)   While this definition can be made and computed with, its meaning and usefulness  remain unclear in the absence of a concrete embedding within string theory.

Another approach, exploiting the relation between the isometry groups and the asymptotic boundaries of Euclidean AdS and de Sitter, is to try to find a bijective map between de Sitter data and EAdS that preserves all the symmetries.     If such a map exists we could map all data on ${\rm dS}_{d+1}$ into ${\rm EAdS}_{d+1}$,  and the Euclidean CFT that is dual to EAdS would also describe de Sitter space.    In fact, such a map preserving the $SO(d+1,1)$ isometry can be constructed:
\begin{eqnarray}
\psi(Y) &=&  \int dX \, G(X,Y) \, \phi(X) \nonumber  \\
G(X,Y) &=& \delta(-X^0Y^0 + X^1 Y^1 +\cdots + X^{d+1} Y^{d+1})
\end{eqnarray}
where $\phi$ and $\psi$ are fields on de Sitter and EAdS respectively.    This map, which is nonlocal by construction, maps the two boundaries of de Sitter into the single EAdS boundary by an antipodal map.   This seems promising, and agrees nicely with the antipodal map in Strominger's proposal, but it transpires that $G(X,Y)$ has a nonzero kernel, so information is lost in the map from de Sitter to EAdS.   This suggests that a single Euclidean field theory cannot be sufficient to describe de Sitter physics holographically.

\subsection{Action, Mass and Time as RG Flow with $\Lambda > 0$: }

Regardless of whether there is a holographic description of de Sitter space, the relation between dS and EAdS asymptopia leads to the solution of a long standing puzzle about asymptotically de Sitter spaces: how do we measure the mass of something placed in such a universe?  In a theory of gravity there is no good local prescription for measuring mass and thus we usually resort to the ADM procedure which  involves surrounding a matter distribution by a large sphere and observing the asymptotic effects on that sphere of the back-reaction on the metric.  In effect, this is a gravitational equivalent of using Gauss' Law to measure electric charges.  Unfortunately, since the spatial sections of de Sitter space are compact, this procedure must also lead to a vanishing mass.   How then do we define and compare the masses of, say, different stars in a closed universe with a positive cosmological constant?

To begin, consider the action (\ref{dsaction1}).   Following \cite{brownyork, hensken, adsmass, skenderismass} we can try to define the mass of an asymptotically dS space by constructing a Euclidean quasilocal stress tensor on either boundary ${\cal I}^{\pm}$
\begin{equation}
\tau_{\mu\nu}  = {2 \sqrt{h}} {\delta S \over \delta h_{\mu\nu}}
\labell{quasilocal}
\end{equation}
and then contracting with the Killing vector $\partial_{t_{II}}$ that is timelike for $r < \ell$ in the static patch (\ref{dsstatic}) and spacelike as $r \to \infty$ at ${\cal I}^{\pm}$:
\begin{equation}
  M = 
    \oint_{\Sigma}  d^{d-1}\phi \,\sqrt{ \sigma } \, N_{\rho} \, 
\epsilon 
    ~~~~~;~~~~~ \epsilon \equiv
    n^{\mu}n^{\nu} \, 
    \tau_{\mu\nu} \, .
\labell{massdef1}
\end{equation}
Here $n^\mu$ is the Killing vector corresponding to $\partial t_{II}$,\footnote{Note that the Killing vector  $\partial t_{II}$ need only exist near the boundaries and it will do so
 at ${\cal I}^\pm$ in any asymptotically de Sitter spacetime.} and we decomposed the boundary metric as
\begin{equation}
    h_{\mu\nu} \, dx^{\mu} \, dx^{\nu } =
       N_{\rho}^{2} \, d\rho^{2} + 
       \sigma_{ab}\, (d\phi^a + N_\Sigma^a \, d\rho) \, 
               (d\phi^b + N_\Sigma^b \, d\rho) \, .
%            N_{\phi}^{2} \, (d\phi + V \, 
%       d\rho)^{2} \, .
       \labell{boundmet}
\end{equation}
The problem, as in AdS space, is that the action (\ref{dsaction1}) diverges because of the large volume at infinity and consequently so does $M$.  In AdS space we understand this in term of the dual field theory -- the effective action and the stress tensor diverge because of the standard local divergences appearing in the definition of composite operators.   In field theory the divergences are removed by adding local counterterms which in turn translates into an improvement of the gravitational action by the addition of local boundary terms that do not affect the equations of motion since they are constructed out of intrinsic  boundary data \cite{adsmass, skenderismass}.   (A useful review is \cite{skenrenorm}.)  

Because of the asymptotic relation between de Sitter space and Euclidean space, the same prescription must work in asymptotically de Sitter spaces independently of whether there is a holographic dual \cite{klemm1,BdBM1,skenrenorm}.  We therefore improve the action (\ref{dsaction1}) by the addition of
\begin{eqnarray}
   I_{{\rm ct}} &=& 
   {1 \over 8\pi G} \, \int_{{\cal I}^{+}} d^{d}x \sqrt{h} \, 
   L_{{\rm ct}}
    +  {1 \over 8\pi G} \, \int_{{\cal I}^{-}}  d^{d}x 
    \sqrt{h} \, L_{{\rm ct}} \labell{counteract} \\
    L_{{\rm ct}} &=&  {(d - 1) \over l} - {l^{2} \over  2(d-2)} R + \cdots
      \labell{counterlag}
\end{eqnarray}
The second term applies when $(d+1)>3$.
The resulting action is finite for $d+1 = 3,4,5$, up to terms related to the conformal anomaly -- a complete analysis of this anomaly and how to treat the associated divergences in both AdS and de Sitter is presented in \cite{hensken, skenderismass, skenrenorm}.   Evaluating the stress tensor on ${\cal I}^-$ gives
\begin{equation}
T^{-\mu \nu} = {2 \over \sqrt{h}}
{ \delta I \over \delta h_{\mu \nu}} = \ \ 
- {1 \over 8\pi G} 
\left[ - K^{\mu\nu} + K \, h^{\mu\nu} - {(d-1) \over l} \, 
h^{\mu\nu} - {l \over (d - 2)} \,G^{\mu\nu}
\right] \, ,
\labell{stressminus}
\end{equation}
where $K$ is the extrinsic curvature and $G^{\mu\nu}$ is the Einstein tensor of the instrinsic boundary metric $h^{\mu\nu}$.

According to this definition, the mass assigned to black holes in de Sitter space is always less than the mass of de Sitter.  In effect, adding matter to the spacetime has causes the spacetime to ``shrink" and the net loss of vacuum energy from $\Lambda$ always exceeds the net increase due to matter.   For example, in three dimensions, adding a pointlike matter source to de Sitter space creates a conical defect spacetime (see e.g, \cite{3ddefects}): 
\begin{equation}
ds^2 = - (m - {r^2 / \ell^2}) \, dt^2 \,  + \,  (m - {r^2 / \ell^2})  \, dr^2 + \,  r^2 \, d\Omega_{d-1}^2
\end{equation}
Here $0 \leq m \leq 1$ with equality for pure de Sitter.  The above formulae lead to
\begin{equation}
M = {m \over 8G}
\end{equation}
so that de Sitter space has the maximum mass in this class.   This result is in attractive correspondence with the fact (\ref{dsentbound})  that empty de Sitter space always has a larger entropy than de Sitter with any combination of matter and black holes in it.    This correspondence suggests a maximum mass theorem for de Sitter space:
\begin{center}
\begin{minipage}[c]{0.9\textwidth}
{\bf Maximum Mass Conjecture}: {\it Any spacetime that is asymptotically de Sitter in the past and has a mass exceeding that of de Sitter develops a cosmological singularity.}
\end{minipage}
\end{center}
This conjecture has passed many tests \cite{mann,moremass}.\footnote{However, recent results of M.~Anderson \cite{anderson} suggest that this conjecture needs to be more tightly specified, for example by restricting the conformal class of the boundary metric.}  An interesting recent paper of Clarkson, Ghezelbash and Mann finds a spacetime which exceeds the de Sitter bound, and does not apparently have a singularity \cite{clarkson}.  However, this space has closed timelike curves.  The classic problem with such spaces is that the boundary of the region of CTCs will be consist of closed null curves along which the stress tensor of any standard field will diverge.  This phenomenon, which is at the core of the Chronology Protection Conjecture \cite{chrono}, will cause the example of \cite{clarkson} to develop a singularity.   It would interesting to directly attempt to prove the maximum mass conjecture.  Since the mass formula is constructed in terms of the extrinsic curvature of the spacetime boundary, standard techniques using the Raychaudhuri equation and focussing theorems might lead to progress.   

If such a maximum mass conjecture is correct in some form, it also points to a route whereby the finite entropy de Sitter space can be consistent with a field theoretic dual -- most initial states would lead to singular evolution 
and the horizon entropy counts the finite number of initial states leading to non-singular spacetimes.   In fact, assuming that 3d de Sitter space is dual to a CFT, and naively applying the Cardy formula using energies and central charges derived from (\ref{stressminus}) gives an exact account of the entropy of the Kerr-de Sitter spacetimes \cite{BdBM1,3ddefects} in three dimensions.    However, some care is required in carrying out such formal procedures, since the unitary representations of the de Sitter algebra are not in standard highest weight representations of the conformal group.   For more on the latter point and possible consequences for a string theory in de Sitter space see \cite{BdBM2,albertodavid}.

\paragraph{Time as RG flow: } Thinking about de Sitter space in the language of the boundary data as described above also leads to a deeper question about the nature of time.   In the AdS/CFT correspondence it was discovered that the physics in the interior of AdS was related to IR dynamics of the dual field theory, while physics near the boundary was related to UV dynamics.   This relation was made precise in Hamiltonian language in \cite{dBVV} and in Wilsonian language in \cite{BK}.  Essentially, radial flow in AdS is related to renormalization group flow in the dual field theory -- the bulk radial coordinate ``emerges'' dyanmically from the field theory as is associated with the field theory RG scale.  If de Sitter space is related to a dual Euclidean field theory we would  expect that RG flow in the field theory would likewise ``generate bulk time'' \cite{BdBM1,stromds2, hologtime}.  In the absence of dual formulation of de Sitter space it is difficult to make this completely precise, but thinking in the way is productive, leading to the identification of a c-theorem and RG-like phenomena (like motion towards a fixed point) in evolution of fields on de Sitter \cite{talltales}.  For example, the contracting phase of de Sitter acts like a flow to the infrared, while the expanding phase acts like a ``reverse'' flow to the UV.   This analogy makes it easier to understand what aspects of mi

\section{Accelerating universes from string theory}

While some progress can be made using symmetries as discussed above, it is essential to find accelerating universes as controlled solutions of string theory.   A major impediment in this regard is a no go theorem that  says, in effect, that no smooth classical compactification of M-theory with lead to an accelerating universe \cite{gibbons,malnun}.  In some ways this theorem is reminiscent of the result that no smooth compactification of M theory will lead to chiral fermions \cite{wittennogo}.     As it turned out compactification of M theory on a classically singular manifold, i.e. an interval, gives the heterotic string with chiral fermions \cite{horavawitten}.  Other escape routes from the no go theorems include the presence of a small internal manifold and thus large $\alpha^\prime$ corrections in string theory, as well as non-perturtabative effects that are absent in the classical Lagrangian.   Of course, if supersymmetry is broken it is essentially inevitable that a positive cosmological constant is generated -- the challenge is keeping it under control.

These loopholes have led to a variety of tactics for constructing accelerating universes in string theory.  Hull has argued that the Type II$^*$ string theories, which result from timelike T-duality, have Euclidean branes the near-horizon limits of which realize de Sitter space \cite{hullds}.   This seems like a tempting scenario in which to realize de Sitter holography, but the challenge is making sense of the Type II$^*$ theories.  The low energy action contains wrong sign kinetic terms which arise originally from the timelike T-duality, an intermediate step involving circular time and therefore closed timelike curves.    However, it is important to realize that there are also higher derivative terms in the action -- it is possible that the combined effect of all of these terms is to give rise to a well-defined theory.  Recent ideas concerning ``ghost condensation'' attempt to implement precisely such a scenario in which higher derivative terms stabilize a theory with ghosts arising from wrong sign kinetic terms \cite{ghostcondense}.  It would be interesting explore how this works in the II$^*$ setting.

Another tactic to evade the no-go theorems forbidding accelerating universes arising from M-theory compactifications is to to explicitly consider classically singular or near-singular compactification manifolds.  By including extra stringy states (e.g. wrapped branes or string that become light near such points it can be possible to construct a smooth low-energy effective action despite the classical singularity (see e.g., \cite{effactsmooth}).    It has been shown that exciting such extra states can lead to periods of accelerated expansion \cite{flopacc} though it is unlikely that inflation can be implemented in this way since the duration of the expansion is typically too short.   Interestingly, the near-singular region of the internal manifold in these cases is dynamically preferred -- once the moduli scalars enter this region, they tend to stay there.

An interesting effort to construct de Sitter vacua of string theory involves asymmetric orientifolds in super-critical string theories \cite{evads}.   These works argue that by turning on fluxes in such models it is possible to stabilize all moduli in a classically reliable region at a positive extremum of the potential.  This results in a metastable de Sitter space.  Traditionally super-critical string theories have been avoided because of runaway behaviours of the dilaton, but in the models of \cite{evads} dilaton tadpoles from non-criticality are balanced against tadpoles coming from the fluxes.  Although the reliability of perturbation theory in such settings is not fully understood, this is a potentially promising approach that shares many features with the more recent constructions of \cite{KKLT}.

 Another interesting approach to finding accelerating universes involves dimensionally reducing string theory on a compact hyperbolic manifold \cite{townsendds}.    One might fear that this would  give rise to wrong sign kinetic terms in the low-energy action on the non-compact space, but there are in fact stable examples in which the non-compact space enjoys a period of acceleration \cite{townsendds}.      The period of acceleration is not long enough to model inflation in the early universe, but this nevertheless an interesting development.\footnote{I thank Joaquim Gomis for emphasizing this work to me.}    Some possibilities for accelerating phases in compactifications on a hyperbolic manifold with flux are classified in  \cite{classif}.    Interestingly, these spacetimes also arise as special cases \cite{ohta} of the S-brane solutions \cite{sbranes} that have been of interest in the context of decaying branes in string theory \cite{sentachyon}.   The latter context might provide a situation where one directly finds a dual description to a time-dependent universe in terms of a Euclidean theory on a decaying brane.   In particular, the ``half-branes'' studied in \cite{halfbrane} represent a soliton that is present at the beginning of time and decays entirely into closed strings \cite{llm}.   One might have hoped that some sort of  decoupling limit of the open-string theory on such decaying branes would give a holographic description of a region of the associated time-dependent spacetimes \cite{sbranes}.    Unfortunately, most attempts to find such spacetimes have led to singular solutions, but very recently, partly inspired by observations in \cite{llm}, non-singular solutions have been found for a class of decaying branes \cite{JAS}.  (Also see the nonsingular S-brane and asymptotically de Sitter solutions found in \cite{nonsingular}.)    The AdS/CFT correspondence was discovered by relating scattering amplitudes from stacks of branes to scattering from the associated spacetime solution.  The techniques for computing scattering amplitudes from decaying branes are under active development (see \cite{llm, schomerus, BKKN}).

Given the idea that the observed cosmological constant and inflation  could arise from the potential energy of a scalar field it is natural to try to use the many light moduli fields generated in typical string compactifications for this purpose.   Typically, however, these fields are difficult to stabilize and when they acquire potentials they tend to be runaways that lead to decompactification to 10 dimensions.  That might be acceptable if the roll rate were slow enough, but this can also be be hard to achieve in traditional perturbative string compactifications \cite{banksdine1}.   

However, recently, it has been realized that many moduli can be stabilized by turning on fluxes on a compactification manifold.   The basic phenomenon, described for example in \cite{polchstrom}, occurs because the pressures generated by the flux threading non-trivial cycles of a compactification manifold make fluctuations in the shape and size of the cycle more energetically expensive.    What is remarkable is that this simple mechanism can freeze a large fraction of the moduli fields in typical compactifications \cite{sethi1,GKP}.    However, at least one modulus (typically the overall internal volume) is left unfixed by the fluxes and in an $N=1$ SUSY compactification this develops a potential that runs away to large volume.   Recently, the authors of \cite{KKLT} (KKLT) realized that non-perturbative effects can  stabilize such a runaway at a supersymmetric extremum.    This basic insight coupled with new techniques for breaking supersymmetry \cite{KPV} led KKLT to propose a generic mechanism for producing metastable de Sitter vacua in string theory.  Below I will review this contruction, point out potential difficulties in constructing concrete models, and then describe the general picture of string vacua that is suggested by the constructions of \cite{KKLT}.\footnote{I am grateful to Per Berglund for the collaboration in the course which I learned the ideas presented below.}

\subsection{Flux vacua and metastable de Sitter}

The proposal of KKLT is phrased in the language of F-theory compactified to 4 dimensions on a 4-fold $X$, or equivalently IIB compactifications on a Calabi-Yau orientifold with 7-branes and fluxes.   This leads to a variety of $N=1$ compactifications with a number of moduli that we seek to freeze.  Of course various dual perspectives are possible but in this setting we can summarize the recipe for constructing metastable de Sitter vacua in four steps:
\begin{enumerate}
\item Ensure consistency by canceling all tadpoles.
\item Stabilize the complex structure moduli classically by using fluxes.
\item Use non-perturbative effects to stabilize the Kahler moduli.
\item Break SUSY in a (more or less) controlled way to get a de Sitter extremum.
\end{enumerate}
String duality will transport this recipe to other pictures of moduli stabilization in different ``duality frames''.

We will study the effective action for 4d, $N=1$ compactifications of IIB string theory on Calabi-Yau orientifolds.  A useful summary of the state of the art in this area is given in \cite{louis} and references therein.  The consistency condition for  F-theory compactifications to 4 dimensions  requires that the net number of D3-branes, and the NS-NS and RR forms be arranged so that
\begin{equation}
{\chi \over 24} = N_{D3} + {1 \over 2 \kappa_{10}^2 T_3} \int_M H_3 \wedge F_3
\labell{consistent}
\end{equation}
where $\chi$ is Euler character of the F-theoory 4-fold or equivalently the charge of the O3 planes and the induced D3 charge on the seven branes in the IIB picture.   $H_3$ and $F_3$ are the NS-NS and RR 3-form fluxes threading the 3-fold base $M$ on which the IIB theory is compactified.   $N_{D3}$ is the net number of space-filling D3-branes located at points on $M$.   Defining
 \begin{equation}
 G_3 = F_3 - \tau H_3
 \end{equation}
 where $\tau$ is the axiodilaton, the Gukov-Vafa-Witten superpotential \cite{gukov} induced by the fluxes is 
 \begin{equation}
 W = \int_M G_3 \wedge \Omega
 \labell{Wdef}
 \end{equation}
 where $\Omega$ is the holomorphic 3-form on $M$.   The Kahler potential is given by a standard  expression of the general form \cite{louis}
 \begin{eqnarray}
 K &=& K_{{\rm k}} + K_{{\rm cs}} ~~~~;~~~~ K_{{\rm cs}} = - \ln \left[ -i \int \Omega \wedge \bar\Omega \right]  \nonumber \\
 K_{{\rm k}} &=& - \ln \left[ -i (\tau - \bar\tau) \right] - 2\ln \left[{\rm Vol}(M) \right] 
 \labell{kahlerpot}
 %\\
% {\rm Vol}(M) &=& \sum_{ijk} a_{ijk}  (\rho^i - \bar{\rho}^i) (\rho^j - \bar{\rho}^j) (\rho^k - \bar{\rho}^k) 
 \end{eqnarray}
 where $K_{{\rm cs}}$ is K\"ahler potential of the complex structure moduli and ${\rm Vol}(M)$ is the volume of the Calabi-Yau, an expression that is cubic in the K\"ahler moduli $\rho^i$ measuring volumes of 2-cycles (i.e. ${\rm Vol}(M) \sim \sum_{ijk} a_{ijk} \rho^i \rho^j \rho^k$).   There are a number of subtle points in determining a good set of K\"ahler cordinates on the moduli space (see \cite{louis} for details), but this schematic form will suffice for the moment.
 % \begin{equation}
% K = -2\ln i[ \sum_{ijk} a_{ijk} (\rho^i - \bar{\rho}^i) (\rho^j - \bar{\rho}^j) (\rho^k - \bar{\rho}^k) + {\rm Compx.~struct.~pieces}]
% \labell{kahlerpot}
% \end{equation}
% where $\rho_i$ are moduli associated to volumes of 2-cycles and the  $a_{ijk}$ are the intersection numbers of the 
% manifold weighted by the appropriate symmetry factors. 
 %(The overall factor of $2$ is absent in KKLT, because they work with a single Kahler modulus and write $K$ in terms of 
 %a 4-cycle modulus, $\rho^\prime \sim \rho^2$.)  
 The standard formula for the potential in $N=1$ 4d supergravity is
 \begin{equation}
  V = e^K \, \sum_{a\bar{b}} {\cal G}^{a\bar{b}}\,  D_a W \, D_{\bar{b}} \bar{W} - 3|W|^2
  \labell{fullpotential}
\end{equation}
where the sum runs over all moduli, ${\cal G}_{a\bar{b}} = \partial_a \partial_{\bar{b}} K$ is the Kahler metric and $D_a = \partial_a + \partial_a(K)$ is the covariant derivative on the moduli space.   Given this structure it is easy to show that V becomes
\begin{equation}
  V = e^K \, \sum_{i\bar{j}} {\cal G}^{i\bar{j}}\,  D_i W \, D_{\bar{j}} \bar{W} 
  \labell{noscale}
\end{equation}
where the sum only runs over the complex structure moduli which in this setting includes the dilaton and the moduli fixing the locations of the 7-branes in the IIB language.  The terms containing derivatives with respect to Kahler moduli cancel with the $3|W|^2$.   As argued in \cite{GKP} a generic choice of fluxes will freeze all the complex structure moduli, but the Kahler moduli, appearing only in the overall $e^K$, will have a runaway behaviour driving them to large values where $M$ decompactifies.

It is important to examine the points at which the complex structure moduli are stabilized and to find the value of the superpotential $W$ at these locations.   (See \cite{klausmirjam1} for a discussion of the location of such supersymmetric extrema in theories with 8 supercharges.)    Supersymmetry demands that $D_a W = 0$ for all the moduli.  This implies that the potential $V =0$ since $D_i W = 0$, so that the Kahler moduli can take any value whatsoever.  However, requiring in addition that $D_\rho W = W \partial_\rho K = 0$ for any Kahler modulus $\rho$ implies that $W = 0$ .   In fact, there is an additional constraint that $G_3$ should be imaginary self-dual ($*_6 G_3 = i G_3$)coming from the requirement that the fluxes be well-defined on the internal manifold \cite{GKP}.  Putting this together with the requirement that $W =0$ in (\ref{Wdef}) implies that at a supersymmetric extremum the complex structure moduli are frozen at a point where $G_3$ is a purely (2,1) form and $W = 0$.   

At this classical level there are also many non-supersymmetric extrema at which $D_i W = 0$ but at which $W = W_0 \neq 0$  so that $D_\rho W \neq 0$.    Such extrema will play a critical role in what follows, because, after the additional of non-perturbative contributions to the superpotential, the non-SUSY extrema can become supersymmetric, stabilizing all the moduli, including the Kahler parameters.  The values of the stabilized Kahler moduli will depend on the value of $W_0$, so it is important to know something about the tree level values that $W_0$ can attain after the complex structure moduli are stabilized.   (We could also consider extrema at which $D_i V = 0$ without necessarily insisting that $D_i W = 0$.  Even after adding the non-perturbative contributions to the superpotential, such extrema will not lead to supersymmetric vacua since they break SUSY in the complex structure directions.)

It has been shown by  Douglas and collaborators \cite{douglascount, douglas2, douglasprivate} that the density of supersymmetric vacua in the $N=1$ moduli space is given by
\begin{equation}
d\mu(z) \propto \det(-R - \omega)
\labell{densityvacua}
\end{equation}
where $R$ and $\omega$ are the curvature and Kahler form of the Kahler metric ${\cal G}_{a \bar{b}}$ on the moduli space.  In essence this is a result stating that the supersymmetric vacua are distributed uniformly in the moduli space.    These techniques can be extended to discuss the distribution of non-SUSY extrema of the potential $V$ as well as the values of the superpotential $W$ at these points but such results have not  been published at the time of this writing \cite{douglas2,douglasprivate}.  Simple explicit examples of the stabilization of complex structure moduli and the value of $W$ at SUSY and non-SUSY extrema have been presented in \cite{examples1,examples2}.

Step (iii) in the KKLT construction is to stabilize the Kahler moduli using non-perturbative effects that depend on the Kahler structure.    The two important effects considered in \cite{KKLT} are:
\begin{enumerate}
\item[(a)] {\bf Brane Instantons}:  Witten \cite{wittinst} argued that Euclidean D3-brane instantons wrapping certain divisors $D$ of the F-theory four-fold $X$, which project to 4-cycles in the base $M$ on which the IIB string theory compactified, will make a contribution to the superpotential.   This will be of the general form $W_{1} = C_1(z) e^{-2\pi {\rm Vol}(D)}$ where $C(z)$ is a one-loop, complex-structure dependent contribution and $ {\rm Vol}(D)$ is the volume of the divisor wrapped by the instanton.  This volume will be a quadratic expression in terms of the 2-cycle moduli $\rho_i$ appearing in the Kahler potential (\ref{kahlerpot}).   Simple examples of this phenomenon are discussed in \cite{wittinst,mayr}.

\item[(b)] {\bf Gaugino condensation: } This effect appears when the complex structure moduli are frozen at a point where 7-branes in the IIB picture coincide leading to an enhanced non-abelian $SU(N)$ gauge symmetry with a gauge couping $1/g^2_{YM} \sim {\rm Vol}(D)$ where $D$ is the 4-cycle in the Calabi-Yau on which the coincident 7-branes are wrapped.  For some examples, see \cite{TT}.   As argued in KKLT, the fact that the complex structure moduli have already been frozen implies that all charged matter (which could be used to separate the branes) has already acquired a mass at a high scale.  So at low energies, the pure $SU(N)$ theory will confine via gaugino condensation and will make a non-perturbative contribution to the   superpotential that behaves as 
$W_2 \sim C_2 e^{-1/g^2_{YM} N} \sim C e^{-{\rm Vol}(D)/N} $.  Again ${\rm Vol}(D)$  will be a quadratic expression in terms of the 2-cycle moduli appearing in the Kahler potential (\ref{kahlerpot}).
\end{enumerate}
To make further progress we must know how many Kahler moduli there are because the effects of the contributions from these two effects may not depend on all of them.

For the moment let us assume  that $h^{1,1}(M) = 1$ so that there is only one Kahler modulus.\footnote{This means of course that $h^{1,1}$ of the F-theory 4-fold $X$ is 2, but one Kahler modulus is frozen in this formulation since the elliptic fiber is shrunk.}   In that case we can set the Kahler potential to be 
\begin{equation}
K = - 2 \ln (i (\rho -\bar{\rho})^3)
\labell{Ksimp}
\end{equation}
(Note that we are using a different convention from KKLT \cite{KKLT} who wrote all their formulae in terms of moduli associated with 4-cycles.  Their Kahler parameter is square of the one appearing in (\ref{Ksimp}).) Then, including the non-perturbative effects described above, the general form of the superpotential is
\begin{equation}
W = W_0 + A e^{a \rho^2}
\labell{Wnonpert}
\end{equation}
where $W_0$ is the tree level superpotential and $A$ and $a$ are some $O(1)$ constants that can depend on the complex structure moduli whose values have already been frozen classically.   Recall that at tree level $W_0 =0$ was required for supersymmetry in order to ensure that $D_\rho W = 0$.   But now, including the non-perturbative effects, the condition for a SUSY vacuum  $D_i W = 0$ gives the condition
\begin{equation}
W_0 = -A e^{-a \sigma^2} (1 + {2 a \sigma^2 \over 2})
\labell{susyext}
\end{equation}
We are setting $\rho = i\sigma$, where sigma is the real volume of the 2-cycle associated to $\rho$.   By solving (\ref{susyext}) we find the stabilized volume of $M$ ($\sim \sigma^3$) in terms of $W_0$.  Also, we can insert the non-perturbatively corrrected superpotential (\ref{Wnonpert}) into the SUGRA formula for the potential (\ref{fullpotential}) and find a potential of the form given in Fig.~2A where we have assumed $W_0 < 0$.   Because $V < 0$, at the single extremum (which is supersymmetric and given by (\ref{susyext})) we have an AdS vacuum.  When $W_0 > 0$ the potential remains a runaway.  Solving to get an extremum in the classically controlled region  of large Kahler modulus it is necessary to fine tune $W_0$ relative to $A$ and $a$ by selecting a suitable set of stabilized complex structure parameters (see \cite{KKLT} for acceptable ranges).

\begin{figure}
  \begin{center}
   \mbox{\epsfbox{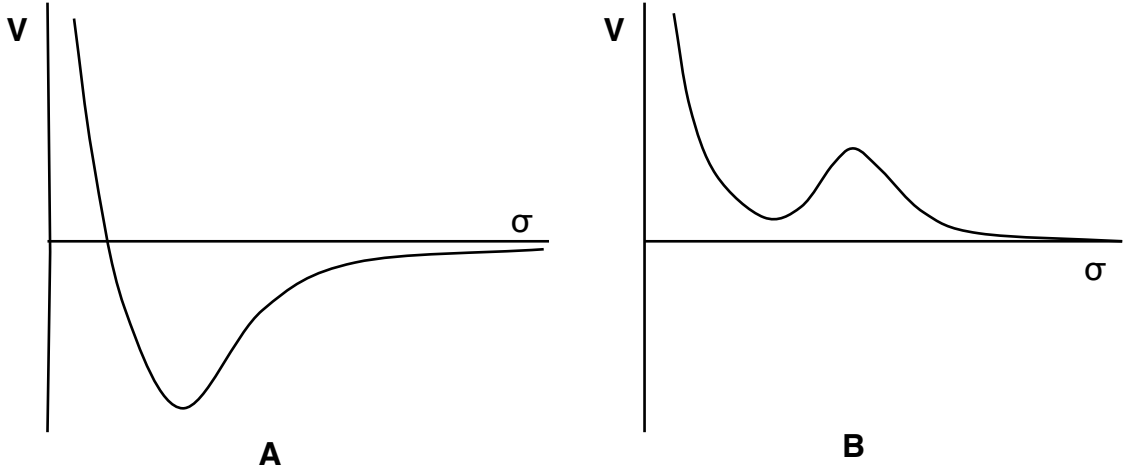}}
\caption{(A) General form for the potential for one Kahler modulus $\sigma$ after including non-perturbative corrections.  The extremum gives a supersymmetric AdS vacuum (B) General form for the potential after breaking SUSY by addition of anti-D3 branes.   The positive local extremum is a metastable de Sitter vacuum which decays by tunneling to the supersymmetric vacuum at large $\sigma$.}
\end{center}
\label{fig:potentials}
\end{figure}

This is already an interesting scenario where all moduli have been frozen.   To get a de Sitter vacuum, supersymmetry will have to be broken.   To do this in a controlled way, KKLT oversaturate the flux contribution to the right hand side of the consistency condition (\ref{consistent}), and then compensate by adding an anti-D3 brane.   One might have thought that an anti-D3 brane added to a scenario where fluxes are creating an effective D3 brane charge would lead to  catastrophic instability.  However, this is not the case \cite{KPV}, essentially because it is difficult for a localized object like a brane to annihilate with a delocalized flux.    In some cases there are perturbative instabilities, but in general this system of an anti-D3 in a background of D3 fluxes decays to the supersymmetric vacuum where the anti-D3 has decayed via a non-perturbative process of bubble nucleation which annihilates one unit of flux.   The lifetime of the metastable state with an anti-D3 in it depends on the details, but can easily be very long \cite{KPV}.  The oversaturated flux and the anti D3-brane make contributions to the potential that go as
\begin{equation}
\delta V = {C \over {\rm Vol}(M)^2} = {C \over \sigma^6}
\labell{deltav}
\end{equation}
where the second equality applies in the single Kahler parameter case with $\rho = i\sigma$.   To understand the $1/{\rm Vol}^2$ scaling note first that the energy of the added flux and that of the anti-D3 will scale in the same way.  The 3-form flux makes a contribution of the form $\int_M G^2$ which involves three inverse factors of the metric in the contraction of the indices of the 3-form $G$.   There is also one integration over $M$.  This gives a net scaling of $ {\rm length}^6 \times ({\rm length}^{-2})^3 = 1$.  Converting from the 4d string frame to the 4d Einstein frame involves a factor of $1/{\rm Vol}(M)^2 \sim 1/\sigma^6$, giving the net scaling of $\delta V$ in (\ref{deltav}).  (Incidentally, the same reasoning is responsible for the overall scaling with $\sigma$ of the potential (\ref{fullpotential}) and the factor of 2 in front of the Kahler potential (\ref{kahlerpot}).   The parameter $C$ depends on the geometry at the location in the Calabi-Yau where the anti-D3 brane is placed.  The term $\delta V$  will lift the potential derived from the nonperturbative superpotential (\ref{Wnonpert}).  Putting everything together, the potential for the single Kahler parameter $\sigma$ is of the form depicted in Fig.~2B, leading to a metastable de Sitter vacuum.  The location and height of the minimum is affected by tuning $W_0$ and $C$ via the choice of stabilized complex structure and the location of the anti-D3 brane.  In particular, this affects the classical reliability of the metastable de Sitter and its stability.  Detailed estimates are given by the authors of \cite{KKLT} under various plausible assumptions.  They argue that it should be possible to tune the fluxes so that the de Sitter extremum has a small cosmological constant (small $V$) and a large lifetime (high barrier).

Thus we have a schematic recipe using the basic of tools standard string theory (Calabi-Yau compactification, fluxes, branes, instantons) to construct metastable de Sitter vacua.   On the one hand the procedure feels contrived because every trick in the book has been used.  On the other, it is very reasonable in the sense that all the standard phenomena we are used to have come into play, and no exotic new phenomena had to be identified.  The urgent question then becomes finding an explicit realization to see whether the proposed scenario can be naturally embedded in an example.

\subsection{The challenge of constructing an explicit example: }

Constructing an explicit example of the scenario described above has proved surprisingly difficult thus far.   General arguments in \cite{GKP} show that a generic choice of fluxes will freeze all the complex structure moduli, but knowing what values of $W_0$ are accessible in this way is harder.   A simple explicit example fixing most of the moduli is given in \cite{examples1} but we really seek a generalization of the arguments in \cite{douglascount,douglas2} that will give us a distribution of values of the stabilized complex structure moduli and associated values  of  the classical superpotential $W_0$ in any given F-theory compactification of interest.

What do the relevant compact manifolds look like?  We are particularly interested in F-theory compactifications to 4 dimensions that can either support the 3-brane instanton or the gaugino condensation effects that lead to Kahler moduli stabilization.   To support the 3-brane instanton, the F-theory 4-fold must be an elliptically fibered Calabi-Yau in which  the 3-brane wraps a divisor $D$ of the base 3-fold $M$.   In order that fermion zero modes cancel giving a nonzero contribution to the superpotential, $D$ must lift to a divisor of $X$ with arithmetic genus 1.  To achieve this it is sufficient for the first three Betti numbers of the lift of $D$ to vanish \cite{wittinst}.  Examples, of F-theory compactifications with the necessary properties are given in \cite{wittinst,mayr,klemm}.  

The simplest relevant 4-fold is an elliptic fibration over a base $B$ which is $P^3$ with a point blown up, leading to IIB theory compactified on $B$ with 7-branes \cite{wittinst}.    This is a model with two Kahler moduli.   Indeed, the 3-brane instanton cannot appear in models with only a single K\"ahler modulus \cite{sethi}.      In order for all K\"{a}hler moduli to be stabilized by 3-brane instantons, there must be a basis for the K\"{a}hler form in which each associated divisor has arithmetic genus 1.      In  typical simple examples with a small number of Kahler moduli the action of the instanton  does not depend on all the Kahler parameters.  As a result, it turns out that there can be a supersymmetric extremum of the nonperturbatively corrected potential, but the extremum occurs on the wall of the Kahler cone \cite{perme}.    Thus, in general the instanton effects might stabilize all the K\"ahler moduli, but they will do so in a region outside classical control, with large stringy corrections.   Nevertheless, Denef, Douglas and Florea \cite{denef} have shown, using results of Grassi \cite{grassi}, that there are many examples of models in which {\it all} K\"ahler moduli are stabilized at reasonably large volumes by instantons.  This explicitly demonstrates the feasibility of the KKLT scenario.    Gaugino condensation effects will enlarge the class of models in which all the moduli are stabilized. Some explicit examples have been studied in \cite{TT}, but all the moduli are not stabilized in these cases.     Further examples will appear in \cite{GKTT}

If supersymmetry is broken by adding an anti-D3 brane as in KKLT, there will be perturbative corrections to the superpotential.   This is the traditional cosmological constant problem in which the corrections to the potential go as $\lambda^4$ where $\lambda$ is the SUSY breaking scale \cite{weinberg2}.  These corrections will certainly increase the value of the cosmological constant and could destabilize the de Sitter extremum in some models.  An important question here is what is the SUSY breaking scale?  On the one hand, if we tune the minimum of the potential in Fig.~2B to get a small cosmological constant, it would appear that there is a small classical breaking scale.  On the other hand, if we look at the amount of energy added in going from the AdS vacuum to the dS vacuum it would appear that the scale is much larger.  This uplift energy comes from the additional fluxes and anti-D3 brane, so we might say that the breaking scale on the added brane is large.   But this will get communicated to the bulk via interactions, so the corrections to the potential will be sizable.  KKLT have checked that within their assumptions these corrections do not destabilize their extremum, but this is an issue that should be paid attention to in constructing such models.  (See \cite{corrections} for two explorations of quantum corrections to the basic KKLT scenario.)  Also, it should also be noted that the K\"ahler potential, even with $N=2$ supersymmetry, receives $\alpha'$ corrections from from perturbative effects and worldsheet instantons \cite{BBHL}.  Orientifolding to get $N=1$ SUSY will increase the class of corrections that appear.   These have the potential to qualitiatively change the structure of the effective scalar potential in some situations \cite{perme}, even possibly leading to SUSY breaking and metastable de Sitter minima.

\subsection{Implications for vacuum selection: the string landscape}
A central problem of string theory is to give an account of why our world looks the way it does given the vast number of apparently consistent vacua with the wrong number of dimensions, too much (super)symmetry, wrong gauge groups and particle content etc.   One might have hoped that non-perturbative considerations would  select four dimensions and low energy particle spectra resembling ours as somehow desirable or unique, but the reverse has happened.  We have shown in great detail, particularly with the understanding of field theoretic and stringy dualities, that the over-symmetric vacua of string theory are self-consistent and cannot be eliminated.   What then selects the vacuum we live in?

The KKLT picture of moduli stabilization and the emergence of de Sitter space suggests a picture of string theory in which there is a potential landscape with many supersymmetric flat regions, non-supersymmetric peaks, valleys and canyons, AdS extrema, metastable de Sitter regions etc.  (See \cite{lennylandscape} for a recent commentary on such a ``landscape" within which the universe evolves.   Some exploration of the landscape occurs in \cite{BdA}.)     The idea of such a landscape goes hand-in-hand with the idea of a {\it discretuum} \cite{boussopolchinski} which states that the less symmetric vacua of string theory relevant to our world are arranged in a dense set of discrete points after moduli stabilization rather than as a well-separated isolated points or on a continuum manifold.   Certainly, this is the indication of the KKLT construction of $N=1$ vacua in the context of generic flux compactifications.  In particular \cite{douglascount,douglas2} have shown that as in Eq.~\ref{densityvacua} the discretely isolated $N=1$ supersymmetric flux vacua of string theory are distributed essentially uniformly within the moduli space.

The potential importance of a discretuum of flux compactifications had been highlighted earlier by Bousso and Polchinski \cite{boussopolchinski} who investigated a modern version of the old proposal of Brown and Teitelboim \cite{brownteit}  that the cosmological constant in 4d could be explained in terms of a space-filling 4-form flux.   The latter authors also proposed that the resulting cosmological constant could dynamically neutralize itself by membrane nucleation.   The necessary flux is naturally obtained in M-theory compactifications to 4-dimensions by compactifying the 7-form flux in 11-dimensions on non-trivial 3-cycles of the internal manifold.  The resulting fluxes can only take quantized values, which initially suggests that the mechanism of \cite{brownteit} will not be viable since the steps in which the cosmological constant can be changed would exceed the current bounds.   However, it is important that in a typical compactification, there will be a great many such 2-cycles in the internal space and hence many different 4-form fluxes in the four noncompact dimensions.  Bousso and Polchinski observed that the net cosmological constant would be
\begin{equation}
\Lambda = - \Lambda_0 + \sum_i c_i N_i^2
\labell{BP}
\end{equation}
where they assumed a large negative bare cosmological constant $-\Lambda_0$ and the $c_i N_i^2$ are the contributions to the ground state energy from $N_i$ quanta of each kind of 4-form.    The key point is that we can find cosmological constants within acceptable bounds  $|\Lambda| < \epsilon$ by locating all points within the $N_i$ charge lattice that lie within a distance $\sim \epsilon$ of a sphere of radius $\Lambda_0$ centered at $N_i = 0$.    It is then elementary to notice that despite the discreteness of $N_i$ there will be a great many lattice points in the acceptable region!   

This observation potentially makes possible an anthropic selection principle for the cosmological constant \cite{weinberg1,boussopolchinski,banksdine3}, since it becomes more reasonable to argue that, given some bound on the total energy density in the system, there is a finite probability to wind up in a vacuum with a cosmological constant like ours.  One might even try to argue for different regions of the universe, separated by domain walls, where the cosmological constant takes different values, and say that we simply live in a region that permits life.   Although the KKLT scenario does not solve the cosmological constant per se, and does not use precisely the mechanism in (\ref{BP}), the presence of a great number of fluxes in the compact manifold raises similar issues concerning a dense discrete set of vacua with similar properties.  It has been argued in this context that vacuum selection in string theory will have to be anthropic -- the best we can do is identify the region of discrete vacua that are phenomenologically acceptable \cite{banksdine2,banksdine3,lennylandscape,douglascount, douglas2}.  All the other vacua are viable too, but they are not seen because they would not give rise to a world like ours.  (The authors of \cite{banksdine2,banksdine3} add some useful cautionary notes concerning the phenomenological import of these ideas.)

While an anthropic explanation of the fundamental parameters of our world is certainly possible, it would be disappointing to many physicists because it suggests a loss of predictability.  In fact, that need not strictly be the case -- it could well be that all viable models in the phenomenologically acceptable region of the discretuum share certain relations between their fundamental parameters, and this would be the prediction of the theory.  In any case, it seems premature to resurrect the anthropic principle because, to date, there is not a {\it single} phenomenologically acceptable model of our world derived from string theory!     Finding one such model may lead to many others, but even one would at present be a triumph.

\subsection{Towards realistic models}

It is worth mentioning that {\it perturbative} effects may suffice to stabilize the Kahler moduli, or at least to give a metastable minimum, in a non-supersymmetric setting \cite{examples2,examples3}.  The basic idea here is that expanding the potential for Kahler moduli around a point with $V > 0$ where the other moduli are stabilized or slowly rolling can modify the no-scale structure of (\ref{noscale}).  Dynamical effects like coupling to the kinetic energy or fluctuations of other fields can also stabilize moduli, at least for a period of time \cite{dynamical}.  These are interesting ideas that point to the fact that many simple dynamical avenues for stabilizing moduli have not been adequately explored.  Likewise, the more exotic avenues towards de Sitter space using II$^*$ theories \cite{hullds,ghostcondense} and non-critical string theory \cite{evads} also deserve exploration.

Using all of these techniques it is interesting to ask what kinds of potentials can be engineered for moduli from string theory.   We know a great many techniques for engineering different low-energy gauge groups and matter content using various compactification manifolds and arrangements of branes.  But to make contact with cosmology we would like to also learn if the kinds of potentials that can give rise to some plausible model of inflation  and perhaps account for the current dark energy can be constructed.   For example, \cite{towardsinflation} found constraints on embedding brane-world inflation into KKLT-like scenarios with stabilized moduli.  Likewise, the authors  of \cite{quevedo} have suggested an interesting model in which the techniques of KKLT \cite{KKLT} give rise to potential with a series of de Sitter minima with cosmological constant decreasing as the compactification scale increases.  This gives a possible stringy realization of the the ``saltatory'' scenarios for the episodic decay of the cosmological constant \cite{brownteit, saltation}.    Several other interesting explorations of inflationary and metastable de Sitter potentials arising from flux stabilization of moduli and supersymmetry breaking via a variety of mechanisms appear in \cite{otherpotentials}.    It would be very interesting to construct a body of rules for engineering moduli potentials with different shapes.

Ultimately we wish to realize inflation and/or a positive cosmological constant in a theory with realistic particle physics.  The two major routes that are used to generate models of low-energy physics with appropriate gauge groups are heterotic compactifications on a Calabi-Yau and compactifications with branes.  In the heterotic setting various groups have established how the vector bundle moduli and the geometric moduli of the compactification manifold can be stabilized using fluxes \cite{heterotic}.  Metastable de Sitter vacua have been argued to arise in \cite{becker,evgeny}.   Fluxes have also been used to stabilize moduli in semi-realistic D-brane  compactifications \cite{realistic}.  In the latter settting there are some tensions between stabilizing moduli while getting chiral fermions in four dimensions, and preserving supersymmetry.  Since a positive cosmological constant breaks SUSY, this tension may not be a bad thing.

Finally, most efforts at constructing models with a small cosmological constant have focussed on first constructing a supersymmetric compactification of string theory and then breaking SUSY in a controlled manner.   Another possibility is to directly consider {\it non-supersymmetric} extrema of the effective potential in the low energy $N=1$ supersymmetric theory generated after adding in the perturbative and  non-perturbative effects that stabilize moduli.  In other words, we would look for extrema of the effective potential $V$ rather than of the superpotential $W$.   In particular, it is possible to obtain {\it non-supersymmetric} AdS minima with a negative cosmological constant.   Since SUSY is broken, there will be a one loop correction to the potential that will lift the extremum towards zero.   Nilles has long speculated that such a mechanism could take an extremum with $V<0$ to a small positive value in a natural way \cite{nilles}.   Unfortunately, detailed studies of such models appear to be in short supply in both the phenomenological and string theory literature.

\section{Conclusion}

Universes with an accelerating expansion are very strange places.   If there really is a positive cosmological constant in our world, there will eventually only be a single quantum within a cosmological horizon.   Nevertheless, Dyson has argued that life might continue in such a universe for a very long time, essentially by having all processes run slower as time passes \cite{dyson}.   Given that our world is not supersymmetric at least below 1 TeV, it should be expected from field theory that a positive cosmological constant is perturbatively generated of a magnitude that is of order  $({\rm TeV}^4)$.  However, the  actual measured value of $\Lambda$ \cite{WMAP,supernovae}  is 120 orders of magnitude smaller than expected from this field theoretic estimate.\footnote{See \cite{weinberg2} for reviews of the cosmological constant problem.}   If it were actually vanishing, we might have hoped that there was some as yet unknown symmetry that guaranteed its absence.  But the apparent non-zero value teaches us that some dynamics is involved.  Understanding the origin of the measured $\Lambda$ and why it can be so small remains one of the outstanding qualitative puzzles of theoretical physics.

\vspace{0.25in}
{\leftline {\bf Acknowledgements}}
This review is an expanded version of lectures given at the Fall 2003 Copenhagen school of the RTN Network and at the PIMS 2003 summer school in Vancouver.  I thank the organizers of both schools for creating very stimulating environments.   I am also  grateful to Per Berglund,  Klaus Behrndt, Jan de Boer, Fawad Hassan, Petr Ho\v{r}ava, Minxin Huang, Esko Keski-Vakkuri, Per Kraus, Albion Lawrence, Tommy Levi, Djordje Minic, Asad Naqvi, Simon Ross and Joan Simon for the many discussions and collaborations in the course of which I learned about the issues described here.   I particularly thank Per Berglund for an ongoing collaboration \cite{perme} concerning the material in Section 4.  This work was supported by the DOE under grant DE-FG02-95ER40893 and by the NSF under grant PHY-0331728.  

%\vfill

%% Bibliography

\vspace{0.25in}
{

\end{document}